\newcommand{\m}{\medbreak}

\newcommand{\no}{\noindent}
\newcommand{\EQ}{\begin{equation}}
\newcommand{\eq}{\end{equation}}
\newcommand{\EQA}{\begin{eqnarray}}
\newcommand{\eqa}{\end{eqnarray}}

\newcommand{\AR}{\renewcommand {\arraystretch}{1.5}
\begin{array}{l}}
\newcommand{\bAR}{\renewcommand {\arraystretch}{2}
\begin{array}{l}}
\newcommand{\ARc}{\renewcommand {\arraystretch}{1.5}
\begin{array}{c}}
\newcommand{\bARc}{\renewcommand {\arraystretch}{2}
\begin{array}{c}}
\newcommand{\ar}{\end{array} \renewcommand {\arraystretch}{1}}

\newcommand{\ET}{\mbox{$E_T\ $}}
\documentstyle[12pt]{article}
\newcommand{\ALLPV}{\mbox{$A_{LL}^{PV}\ $}}

\newcommand{\r}{\rightarrow}

\newcommand{\Z}{$Z^{\circ}\ $}
\newcommand{\ZP}{$Z'\ $}
\newcommand{\WP}{$W'\ $}

\begin{document}
\begin{titlepage}
\vspace{0.2in}
\vspace*{1.5cm}
\begin{center}
{\large \bf Sensitivity to a new right-handed charged  current in polarized hadronic
collisions at RHIC
\\} 
\vspace*{0.8cm}
{\bf P. Taxil} and {\bf J.M. Virey}{$^1$}  \\ \vspace*{1cm}
Centre de Physique Th\'eorique$^{\ast}$, C.N.R.S. - Luminy,
Case 907\\
F-13288 Marseille Cedex 9, France\\ \vspace*{0.2cm}
and \\ \vspace*{0.2cm}
Universit\'e de Provence, Marseille, France\\
\vspace*{1.8cm}
{\bf Abstract \\}
\end{center}
In the context of general
$SU(2)_L\otimes SU(2)_R\otimes  U(1)\;$ models a relatively light
\WP is still allowed in some scenarios with a heavy right-handed neutrino
and a quark mixing matrix $U^R$ close to unity.
We explore the consequences of the presence of this new charged current
on the parity violating spin asymmetries which could be induced in
one-jet inclusive production in polarized proton-proton or proton-neutron
collisions. Such measurements could be performed within a few years
at the Brookhaven Relativistic Heavy Ion Collider (RHIC) running part of 
the time as a polarized hadronic collider.

\vfill
\begin{flushleft}
\m\no
Number of figures : 2\\

\m\no
April 1997\\
CPT-97/P.3469\\
\m\no
anonymous ftp or gopher : cpt.univ-mrs.fr

------------------------------------\\
$^{\ast}$Unit\'e Propre de Recherche 7061

{$^1$} Moniteur CIES and allocataire MESR \\
E-mail : Taxil@cpt.univ-mrs.fr ; Virey@cpt.univ-mrs.fr
\end{flushleft}
\end{titlepage}

\section{Introduction}
\indent
\m
The possibility of
new charged current interactions for the right-handed fermions
has been considered for a long time as one of the direct extensions
of the electroweak Standard Model (SM).
The simplest example is the left-right symmetric model based on the
gauge group
$SU(2)_L\otimes SU(2)_R\otimes  U(1)_{B-L}\;$  \cite{PatiSalam}
which implies the presence of three new gauge bosons, two charged $W'{^{\pm}}$
and one neutral \ZP.
\m
It is well known that the new charged bosonic sector is more severely bounded 
by low-energy data than the neutral one. 
The most severe constraints are coming from the 
$K_S - K_L$ mass difference $\Delta m_K$. These constraints, however,
rely on some specific assumptions (for a nice review see \cite{LangackerSankar}
and references therein).   
\m
For the symmetric models with $g_L = g_R$, and if one assumes an equal magnitude for
the left-handed and right-handed quark mixing matrix elements $U^L_{ij}$ and
$U^R_{ij}$ ("manifest L-R symmetry"), the bound which can be obtained on the
mass of the heavy right-handed charged boson is rather stringent : 
$M_{W'} \geq 1.4 - 2.5\;$ TeV, with the exact value depending on the values of the QCD
enhancement factor \cite{BBS}.
The limits are the same for "pseudo-manifest
L-R symmetry" that is  if  $\mid U^L\mid = \mid U^R\mid$. 
Note that the approximations in the estimates of the hadronic matrix elements
introduce some uncertainty : an approach based on the QCD sum rules lowers this bound
down to 700 GeV \cite{Colangelo}.
In the same time the $W$-\WP mixing angle is also severely bounded from
universality :  $\mid \xi \mid < 0.003$. 
However, if more freedom is allowed on the gauge couplings
("general left-right models" with $g_L \neq g_R$)
and especially on the values of the matrix elements $U^R_{ij}$, most of the constraints
can be evaded. The weakest (90\% C.L.)
bound for $M_{W'}$ advocated in  \cite{LangackerSankar} 
correspond to $U^R \approx I$ (identity), it corresponds to : 
\EQ 
M_{W'}/\kappa \geq 300\, {\rm GeV}
\eq
\no
where $\kappa = g_R/g_L$. Concerning the mixing angle $\xi$ one still gets a bound in
the $1\%$ range:  $\kappa \mid \xi \mid < 0.013$. 
Following grand unification arguments, it is usually assumed that
$0.55 \leq \kappa \leq 1$ which can give us a \WP as light as $\approx$ 170 GeV.
Note that, with the present value of the top mass, these relatively old limits can be
lowered by at least 10\% \cite{LangackerSankar,RizzoWright}.
\m
Concerning direct searches, in ${\bar p p}$ collisions, with the hypothesis that the
$\nu_R$ is sufficiently light to allow the decay \WP $\r l_R\, \nu_R$, CDF obtained
the  limit : $M_{W'} > 652\,$ GeV (at 95\% C.L.) \cite{CDFWplepton}. This limit is
obtained in the leptonic channel in case of $\kappa = 1$ and $U^R = U^L$. 
From low energy muon decay analysis, with the same assumptions, the bound is 406 GeV
\cite{PDG96}. On the other hand, if the \WP leptonic decay is forbidden, due to a
very large mass value for the $\nu_R$, the constraints are much weaker since the direct
search in hadronic collisions is then restricted to the two-jet decay channel
(the $WZ$ decay channel will also be promising but only at LHC \cite{WZchannel}). 
In this situation, the mass
domain excluded by UA2  \cite{UA2Wright}
$100 \, < M_{W'}\,  < 250\,$ GeV (at 90\% C.L.) has been very recently extended by CDF
\cite{CDFWjets2}:  $300 \, < M_{W'}\,  < 420\,$ GeV (at 95\% C.L.).
Both experiments assumed $\kappa = 1$. From CDF published data, one can infer that the
lower bound on \WP is disappearing if the value of $\kappa$ is reduced
below $\kappa = 0.95$.

Actually, due to the peculiarities of the two-jet channel, a window is still open in
the region below 100 GeV. This region is not strictly forbidden 
neither by $\Delta m_K$
nor by bounds from double-beta decay as soon as $\, i)$ the right-handed neutrino is a
massive Dirac neutrino $\, ii)$ extreme fine tuning of the $U^R$ parameters 
(three angles and six phases) is allowed  to get a cancellation between terms
involved in the calculation of $\Delta m_K$ \cite{LangackerSankar}. 
The existence of this window has been noticed at the recent HERA workshop
\cite{HERAworkshop} : at HERA, in this case, high luminosities should allow a
significant discovery potential.   \m
Many studies have been
devoted to the sensitivity of the ongoing Tevatron searches for the high \WP mass
domain,
assuming that the missing $E_T$ leptonic decay mode is accessible and dominant
\cite{RizzoWright,RizzoSnowmass}. 
On the other hand, direct detection of the \WP in the two-jet channel will probably
remain difficult and full of the uncertainties inherent to this channel.
\m
From now, we will place ourselves in the scenario where a very massive right-handed
neutrino forbids the leptonic decay. As discussed above, in the case where  
$U^R \approx I$, a relatively light \WP is still allowed, with various
possible windows for the values of its masses, depending on
the value of $\kappa$. 

In fact, the presence of  such \WP exchanges in the quark-(anti)quark
scattering subprocesses could induce some 
other kinds of deviations from the SM expectations, at
sub-energies $\sqrt{ \hat s}$ below $M_{W'}$. Due to the
right-handed structure of the new current, it should give rise to particular 
parity violating (PV) spin effects and this should be exploited. Since, in the context
of hadronic collisions it is quite hopeless to analyze the net helicity of an outgoing
particle or jet,  polarized beams are needed to build an initial-state spin asymmetry. 
Indeed, we have shown recently that, in polarized $pp$ collisions, the presence of a
hadrophilic \ZP \cite{TVZprime}
and/or a new PV contact interaction between quarks \cite{TVCT2}
could yield some
deviations from the spin asymmetry in one-jet inclusive production
which is due to SM electroweak (EW) bosons exchanges. 

Theses analyses were performed in the context the Relativistic Heavy Ion Collider
(RHIC) which will be run part of the time as a polarized hadronic collider. This program
\cite{BuncePw} will start in a few years from now, with high intensity polarized proton
beams and a center of mass energy from 200 GeV (for the first run)  up to 500-600
GeV. The possibility of accelerating polarized  $^3He$ nuclei, which has been discussed
recently \cite{Amsterdam,RSCmeeting},
will open some new perspectives since polarized $pn$ collisions will be allowed.  

\m  
In the following we will discuss the influence of the new \WP on the one-jet PV spin
asymmetry \ALLPV in polarized $pp$ and $pn$  collisions at the planned RHIC energies
and luminosities.  We first present the ingredients entering into the calculation of
the asymmetry \ALLPV, then we give the limits which could be obtained on the parameter
space ($\kappa,M_{W'}$) of the new right-handed charged sector, given the sensitivity
of the polarized RHIC experiments. 
\m
\section{Double helicity PV asymmetry in one-jet inclusive
production}

For the inclusive process  $H_a\ H_b \ \r jet + X$,
when both beams can be polarized (this is the case at RHIC), one  defines a
double helicity PV asymmetry : 
\EQ
\label{ALLPVdef}
A_{LL}^{PV} ={d\sigma_{a(-)b(-)}-d\sigma_{a(+)b(+)}\over 
d\sigma_{a(-)b(-)}+d\sigma_{a(+)b(+)}}
\eq
\noindent
where the signs $\pm$ refer to the helicities of the colliding hadrons.
From now, $d\sigma_{a(h_a)b(h_b)}$ will mean the
cross section in a given helicity configuration $(h_a,h_b)$,
for the production of a single jet
at a given transverse energy $E_T$ and pseudorapidity $\eta$ :
\EQ
\label{dsrap0}
d\sigma_{a(h_a)b(h_b)} 
\; \equiv \; {d^2\sigma^{(h_a)(h_b)} \over {dE_T d\eta}}
\eq
\no
In the following, we integrate $d\sigma$ over a pseudorapidity interval
$\Delta \eta =2.6$ or 1.0 (see below)  centered at $\eta=0$, and over an $E_T$ bin
which corresponds to a jet energy resolution of 10\%. 
\m
Any helicity dependent hadronic cross section is obtained by convoluting appropriately
the subprocess cross sections 
${d\hat \sigma}_{ij}^{\lambda_1,\lambda_2}/d {\hat t}\;$, which depend upon the
parton helicities $\lambda_1$ and $\lambda_2$, with the polarized quark and/or
antiquark distributions  evaluated at some scale $\mu^2$:
$q_{i\pm}(x,\mu^2)$ and $\bar q_{i\pm}(x,\mu^2)$ (explicit formulas can be found in
\cite{BRST,BouGuiSof}).
Here, $q_{i\pm}$ means the distribution of the polarized quark of flavor $i$ having its
helicity parallel (+) or antiparallel (-) to the parent hadron helicity.
It is usual to define $\Delta q_i = q_{i+} - q_{i-}$.
The chosen $\mu^2$ value is $\mu^2 = E_T^2$, we have checked that changing this 
choice has no visible influence on our results. 
 
We follow the notations of \cite{BouGuiSof} where :
\EQ
{d\hat \sigma_{ij}^{\lambda_1,\lambda_2}\over d\hat t} 
\; =\;{\pi\over \hat s^2} \,
\sum_{\alpha,\beta} T_{\alpha,\beta}^{\lambda_1,\lambda_2}(i,j)
\eq
\no
$T_{\alpha,\beta}^{\lambda_1,\lambda_2}(i,j)$ denoting the matrix element squared with 
$\alpha$ boson and $\beta$ boson exchanges, in a given helicity configuration for
the involved partons $i$ and $j$ . 
\m
QCD dominates the unpolarized cross section and it is not difficult to incorporate
into the calculation the tiny EW terms, their interference with QCD amplitudes 
\cite{AbudBaurGloverMartin} and also
the Non Standard \WP terms with a \WP coupling to quarks of the form
\EQ
\imath {g_R\over \sqrt 2} \bar q_i {1\over 2} (1+\gamma_5) \gamma_{\mu}
W'^{\mu}\, U_{ij}^R q_j
\eq
These latter have a small influence on the unpolarized cross section in the range of
parameters we consider. At RHIC energies it is then hopeless to isolate a bump in the
dijet mass spectrum if $M_{W'}$ lies in the range which is still allowed. 

Concerning \ALLPV, the leading order (LO) SM contribution 
is known for a long time \cite{PaigeRanft}. It comes essentially from interferences
between QCD and standard EW PV amplitudes (see \cite{BouGuiSof}
for the correct expressions for the relevant $T_{\alpha,\beta}$'s). It is raising with
$E_T$ at large $E_T$ due to the increasing influence of $qq$ scattering
\cite{TannenbaumPenn} and it was reestimated recently using modern $\Delta q_i$'s in
\cite{TVCT2,TVZprime} for $\vec p \vec p$ scattering.

Note that it can be advocated that a LO calculation of \ALLPV gives
a quite good estimate since, QCD being helicity conserving in the limit of massless
quarks, one does not expect a significant influence of NLO corrections on the spin
asymmetries. Indeed, when NLO calculations are available as is the case for
inclusive prompt photon production, it has been stressed that the 
spin asymmetries were less influenced by the NLO corrections than the
individual polarized or unpolarized cross sections \cite{Gordon2}. 
\m 
Let us now concentrate on the effect of the new right-handed current.\\
In the \ET range of interest the
contribution of quark-quark  scattering $q_iq_j\,(i\neq j)$ dominates over terms
involving antiquarks (these latter are carefully taken into account in the full
calculation).  In short notations \ALLPV is then given by the expression :  
\EQ
\label{ALLPVjet}
A_{LL}^{PV} . d\sigma \simeq - \sum_{i,j} 
\int
T_{g,W'}^{++}(i,j)
\biggl[q_i(x_1,\mu^2)\Delta q_j(x_2,\mu^2) + \Delta q_i(x_1,\mu^2)q_j(x_2,\mu^2) 
+ (i\leftrightarrow j) \biggl]
\eq \no
where $T_{g,W'}^{++}(i,j)$ is the term originating from the
interference between one gluon exchange and the \WP exchange amplitudes
for right-handed quarks 
(taking care of color rules and crossing symmetry) :
\EQ
T_{g,W'}^{++}(i,j) = {4\over 9} \alpha_s (\mu^2)
{g_R^2 \over \pi} {|U_{ij}^R |}^2 \, {{\hat s}^2 \over {\hat u_{W'} \hat t}}
\eq
where $\hat u_{W'} = \hat u - M^2_{W'}$.
\m
Since the scattering of valence quarks of different flavours is the dominant process,
it is easy to understand why proton-neutron collisions will be preferred to the more
familiar proton-proton case.

\m
For ${\vec p}\,{\vec p}$ collisions, 
when charged bosons $W$ (or \WP) are involved 
and if one neglects the heavy flavour content of the
nucleon, only the expressions $u\Delta d + d\Delta u$ will enter in
eq.(\ref{ALLPVjet}). 
Some very general tendencies of the polarized quark distributions are : \linebreak 
$\, i)$ $\Delta u >
0$, $\Delta d < 0$ ; $\, ii)$ $u\Delta u \gg d|\Delta d|$ ;  $\, iii)$ $ \Delta u/u >
|\Delta d|/d$. 
Therefore, in case of $W$ (\WP) exchanges, there is a partial
cancellation in the $u\Delta d + d\Delta u$ term. This cancellation is much less
important in case of \Z (or neutral \ZP) exchanges 
where the combination $u\Delta u + d\Delta d$ dominates in the
corresponding formula (see \cite{TVZprime}) thanks to the dominance $\,ii)$.  Note that
this partial cancellation is somewhat compensated by the maximal PV in
eq.(\ref{ALLPVjet}) compared to the relatively smaller amount of PV which is present in
the neutral sector (at least when the standard \Z is involved).\\ 
In ${\vec p}\,{\vec n}$ collisions the situation is reversed if one invokes Isospin
symmetry which implies $u^n = d^p(\equiv d), \, d^n = u^p(\equiv u)$ and
$\Delta u^n = \Delta d^p, \, \Delta d^n = \Delta u^p$ in the same way. The net
results is that ${\vec p}\,{\vec p}$ collisions are well suited to see the effect of 
gluon-neutral gauge boson (\Z and/or \ZP) interferences, whereas in 
${\vec p}\,{\vec n}$ collisions, the interferences between gluons and charged gauge
bosons ($W$ and/or \WP) are enhanced.

In the following, we will only present the results of our analysis on the most
interesting channel of polarized ${\vec p}\,{\vec n}$ collisions.
\section{Discussion and results}
 In Fig.1 we show the asymmetry \ALLPV in polarized ${\vec p}\,{\vec n}$ collisions
versus the transverse jet energy $E_T$ at a c.m. energy $\sqrt s \approx 500 GeV$ that
is for a 300 GeV proton beam colliding with a 600 GeV 
beam of polarized $^3He$ nuclei.
We have chosen for
illustration  $M_{W'}\,$ = 100 GeV, 300 GeV and 400 GeV (for $\kappa = 1$) and
200 GeV (for $\kappa = 0.8$). 
The standard \ALLPV,  which is bigger than the corresponding one in 
${\vec p}\,{\vec p}$ collisions, is shown for comparison. 
The error bars correspond to
the statistical error for a degree of polarization $\cal P\,$ = 70\% 
(see \cite{TVZprime,TVCT2}) and an
integrated luminosity of 800 $pb^{-1}$ which can be achieved in a few months running,
the systematic error being negligible \cite{RSCmeeting}. We have chosen an
$E_T$ region where these errors are small : the range between 60 GeV and 100 Gev will
dominate the analysis which follows.

This calculation has
been performed for $U^R = I$. In fact we are not sensitive to the precise form of $U^R$
as long as  the off-diagonal matrix elements are small. Since the $W$-$W'$ mixing angle
is already severely restricted  (less than $\approx$ 1\%) we can neglect
safely its effects. These results are also independent of the precise value 
of $m_{\nu_R}$ as soon as one remembers that we placed ourselves in the situation where
the decay  $W' \r \nu_R l_R$ is forbidden by kinematic. 
We have used for consistency
LO spin dependent distributions, namely the ones of \linebreak GRV \cite{GRV} which fit
well the polarized deep-inelastic scattering data. It has to be remembered that our
predictions are not affected by the present uncertainties on the polarized gluon
distributions. The first part of the polarized RHIC program itself
\cite{BuncePw,RSCmeeting} will greatly improve our knowledge of $\Delta q_i$'s and
$\Delta \bar q_i$'s.

One can see from Fig.1 that the measurement of \ALLPV at RHIC should allow to pin
down easily the presence of a right-handed \WP with a mass around or below
100 GeV. In this case \ALLPV is compatible with zero because an important cancellation
occurs between the standard $W$-gluon and the \WP-gluon interference terms. 
A zero
\ALLPV is clearly forbidden in the SM, for any reasonable choice for the spin
dependent quark distributions. With the high precision achievable at RHIC, thanks
to the high luminosity, such an effect cannot be missed.

\m
We present in Fig.2 the limits we obtain at RHIC on the parameter space ($\kappa,
M_{W'}$). The shaded areas correspond to the two zones excluded respectively  by UA2
and CDF in the two-jet channel. From the published results we have smoothly 
extrapolated their limits down to values of $\kappa$ below 1. The dashed line
corresponds to the "theoretical" 90\% C.L.
upper bound $M_{W'} = 300 \, \kappa$ (in GeV) 
according to eq. (1). Remember that this bound can be avoided if extreme fine tuning
of the $U^R$ matrix elements is allowed as discussed in \cite{LangackerSankar}.
The two other lines correspond to our 95\% C.L. upper limits
which can be obtained at RHIC (with the same figures for the luminosity and the energy
as above) after having integrated over a pseudorapidity domain
$\Delta \eta = 1$ or $\Delta \eta = 2.6$. The first case corresponds to a "minimal"
detector configuration, the second one to the case where it will be possible to
extend the rapidity interval by adding "end-caps" to the STAR detector at 
RHIC \cite{RSCmeeting}.

These bounds display an approximate scaling : 
$M_{W'} = 400 \,\kappa - 10$ (in GeV) for $\Delta \eta = 1$ and
$M_{W'} = 500 \,\kappa - 30$ (in GeV) for $\Delta \eta = 2.6$.
It is clear that the second option is preferred since it allows to probe the presence
of a new \WP up to $M_{W'}$ = 470 GeV (for $\kappa =1$) instead of
390 GeV, a value already excluded by the CDF bound.
\m
One can see that, if $\kappa$=1, there is little room for discovery except in the
high mass region between 420 GeV and 470 GeV, in the narrow window between 250 GeV
and 300 GeV and in the region below 100 GeV as discussed above (ignoring the bound
eq. (1) in these last two cases). On the other hand, the precise measurement of
\ALLPV should allow to cover the "small $\kappa$" region  ($0.55 < \kappa <1$) for
relatively light \WP. For large $\kappa$ values, it is possible to probe
$M_{W'}$ up to 670 GeV, with a detector with maximal coverage.
In the case of ${\vec p}\,{\vec p}$ collisions, the bounds from \ALLPV 
fall into the region already excluded by UA2. The region $M_{W'} \leq$ 100 GeV is
of course still interesting in this configuration.

\section{Conclusion} 
A consequence of the scenario with a very massive right-handed neutrino is the
difficulty to pin down the presence  of a new right-handed charged gauge boson \WP
which is supposed to decay only into a pair of jets with no missing energy. 
Looking in this channel CDF
and UA2 had left open some consequent windows in the parameter space.
On the other hand it has been stressed that low energy bounds 
on $M_{W'}$ can be evaded, leaving open the possible existence
of a light or relatively light \WP.

Polarized hadronic collisions offer a
unique opportunity to get a handle on  the interference effects which could be induced
by the new charged current.  Indeed, the right-handed nature of the \WP introduces some
disturbance on the spin asymmetry \ALLPV in inclusive one-jet production. These
effects could be seen at RHIC for  quite a large portion of the parameter space
($\kappa,M_{W'}$).  Of course, a great precision is needed :
at RHIC this can
be achieved in a few months running thanks to the very high luminosity \linebreak 
(${\cal L} = 2.10^{32} cm^{-2}s^{-1}$) and to the high degree of beam polarization.
To complete this program, the availability of polarized neutrons is necessary and this
can be obtained with beams of $^3He$ or Deuteron nuclei.

Finally, if a
remarkable deviation from the standard \ALLPV  is observed, then it will be
mandatory to perform some careful analysis to get some informations on the true nature
of the new interaction. Disentangling the effect of a hadrophilic \ZP from the one of a
new \WP or from a flavour conserving and parity violating Contact Term will not be an
easy task. For this purpose, having at disposition both polarized proton and neutron
beams could greatly help. These questions will be treated in more detail in a
forthcoming analysis.


\newpage
\no {\bf Acknowledgments}

\m \no
We thank J. Soffer for fruitful discussions, A. Zylberstejn for information on the FNAL
experiments and the members of the RHIC Spin Collaboration (RSC) for useful
discussions at the occasion of the recent Marseille's RSC meeting.



\newpage
{\bf Figure captions}
\bigbreak
\no
{\bf Fig. 1} \ALLPV for one-jet inclusive production, versus $E_T$, for polarized
$p\,n$ collisions at RHIC at a c.m. energy of 500 GeV. Standard model expectation
(plain curve), \WP effects with a \WP mass of 400, 300, 100 GeV ($\kappa =1$)
and 200 GeV ($\kappa = 0.8$). The error bar corresponds to the statistical
error (with an integrated luminosity of 800 $pb^{-1}$). The rapidity interval is 
$\Delta \eta = 2.6$.

\bigbreak \no
{\bf Fig. 2} Bounds on the parameter space ($\kappa,M_{W'}$) in polarized $p\, n$
collisions. The shaded areas correspond respectively to the
CDF and UA2 excluded regions (in the two-jet channel). The dashed curve is the 
theoretical upper bound eq. (1), the plain (dotted) curve  corresponds to the RHIC
upper limit from \ALLPV with the same collider parameters as in Fig.1 
with $\Delta \eta =$2.6 ($\Delta \eta =$1.0).

\bigbreak
\no

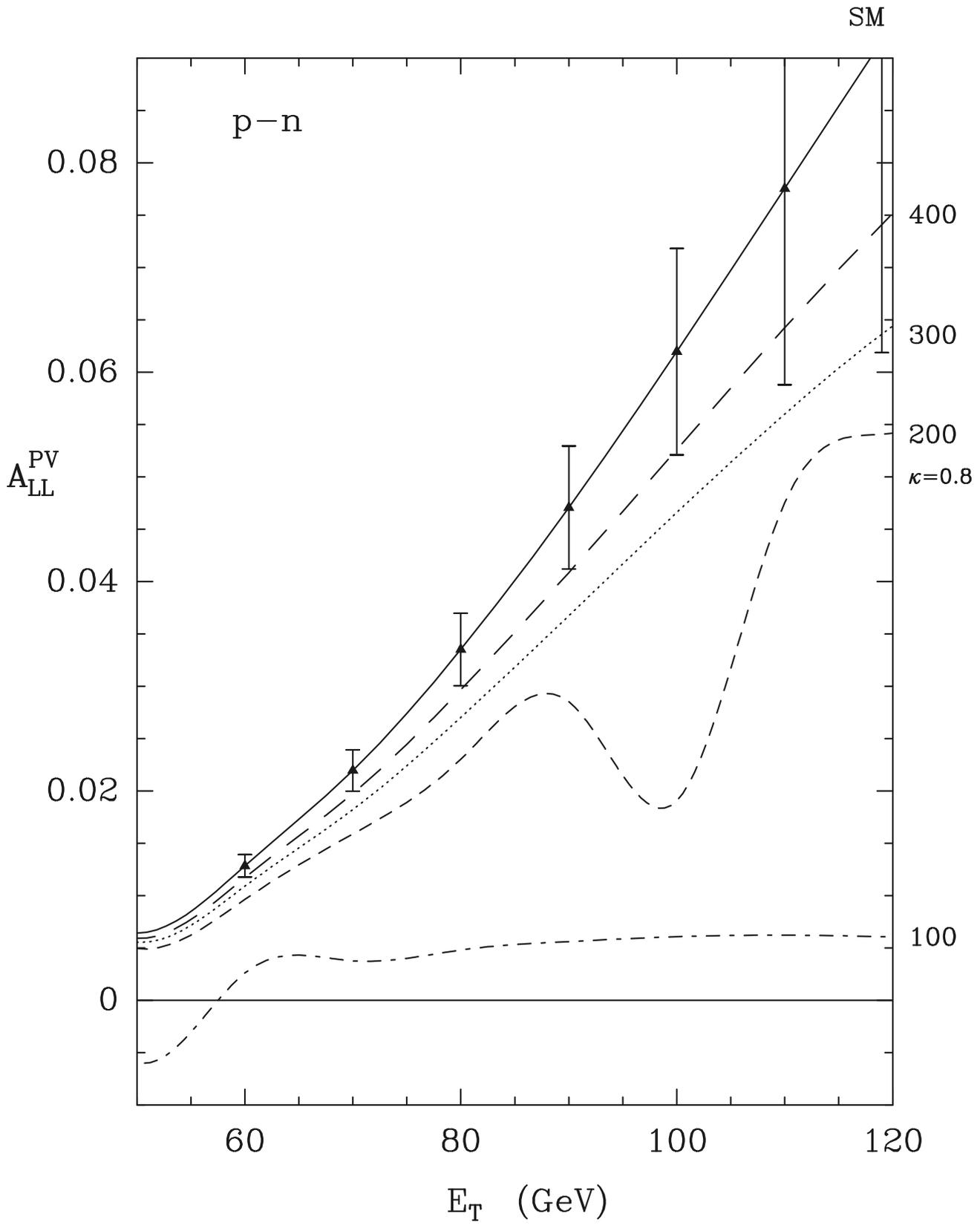
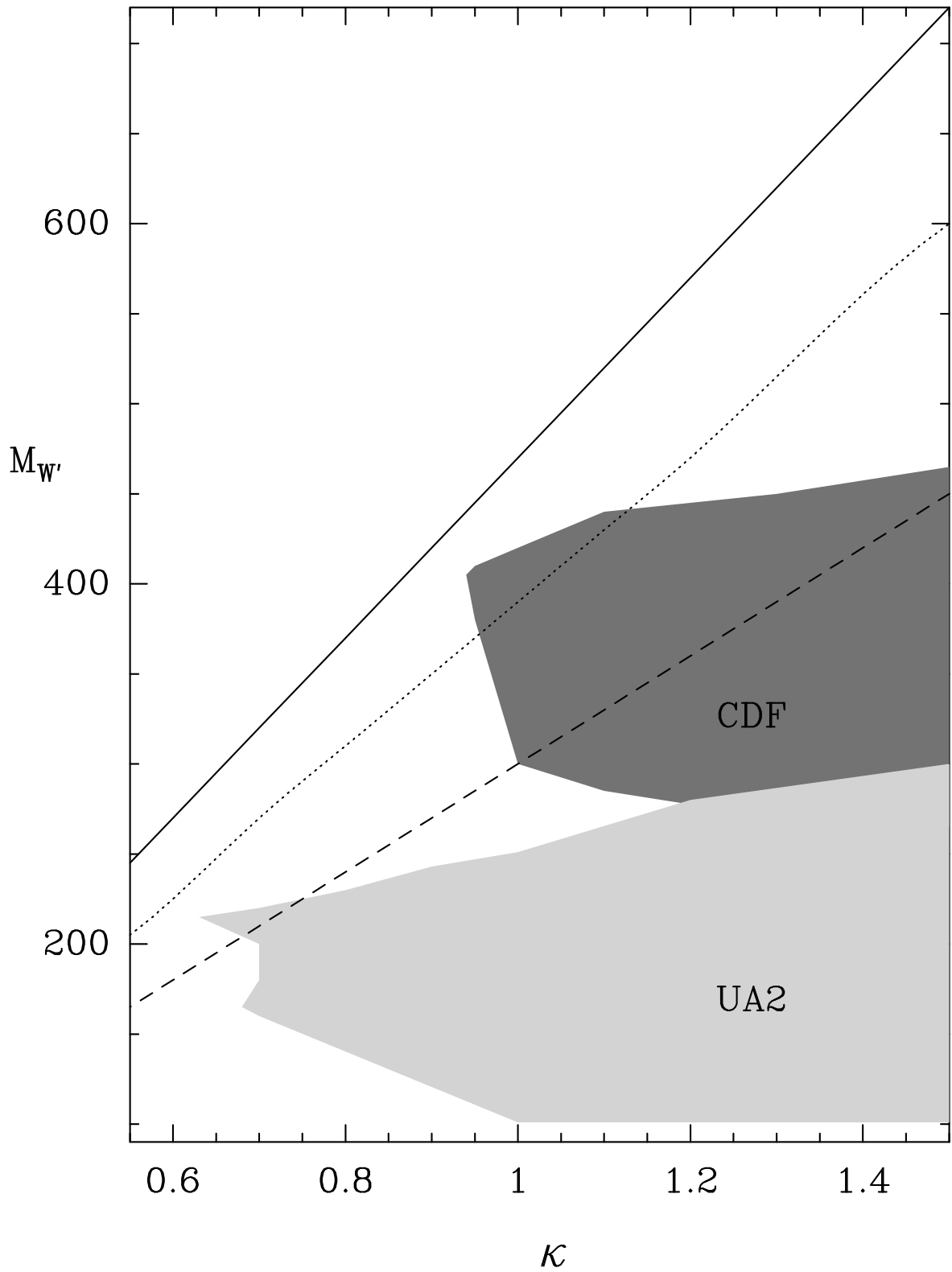
\end{document}